\documentclass[aps,prb,preprint,showpacs,superscriptaddress]{revtex4-1}
\usepackage{graphicx}
\usepackage{amsmath}
\usepackage{amssymb}
\usepackage{color}
\usepackage{epstopdf}
\usepackage{float}
\usepackage{multirow}

\newcommand{\lux}{\affiliation{Physics and Materials Science Research Unit,
                               University of Luxembourg,
                               162a avenue de la Fa\"{i}encerie,
                               L-1511 Luxembourg, Luxembourg}}

\begin{document}
\title{Inter- and intra-layer excitons in 
MoS$_2$/WS$_2$ and MoSe$_2$/WSe$_2$ heterobilayers}

\author{Engin Torun} 
\email{engin.torun@uni.lu}
\lux

\author{Alejandro Molina-S\'{a}nchez}
\lux
\affiliation{
Institute of Materials Science (ICMUV),
             University of Valencia, Spain}

\author{Henrique P. C. Miranda}
\lux
\affiliation{Institute of Condensed Matter and Nanosciences (IMCN/NAPS),
             Universit\`{e} catholique de Louvain, Belgium}

\author{Ludger Wirtz} 
\lux

\date{\today}

\begin{abstract}
Accurately described excitonic properties of transition metal dichalcogenide heterobilayers (HBLs) are crucial to
comprehend the optical response and the charge carrier dynamics of them. Excitons in multilayer systems posses inter or intralayer character whose spectral
positions depend on their binding energy and the band alignment of the constituent single-layers. In this study, we report the electronic structure
and the absorption spectra of MoS$_2$/WS$_2$ and MoSe$_2$/WSe$_2$ HBLs from first-principles calculations. We explore the spectral positions, binding energies and the origins
of inter and intralayer excitons and compare our results with experimental
observations. The absorption spectra of the systems are obtained by solving the
Bethe-Salpeter equation on top of a
G$_0$W$_0$ calculation which corrects the independent particle eigenvalues obtained from density functional theory calculations. Our calculations reveal that
the lowest energy exciton in both HBLs possesses interlayer character which is decisive regarding their possible device applications.
Due to the spatially separated nature of the charge carriers, the binding energy
of inter-layer excitons might be expected to be considerably smaller than
that of intra-layer ones. However, according to our calculations the binding energy of lowest energy interlayer excitons is only $\sim$ 20\% lower due to the weaker screening of the Coulomb interaction between layers of the HBLs.
Therefore, it can be deduced that the spectral positions of the interlayer excitons with respect to intralayer ones are mostly determined by the band offset of the constituent single-layers.
By comparing oscillator strengths and thermal occupation factors, we show that
in luminescence at low temperature, the interlayer exciton peak becomes
dominant, while in absorption it is almost invisible.
\end{abstract}

\maketitle

\section{Introduction}
Single-layer transition metal dichalcogenides (TMDs) are paradigmatic materials due to their strong light-matter
interaction, and remarkable excitonic effects on the optical properties.\cite{Mak2010,Splendiani2010,Britnell2013,Lopez-Sanchez2013} 
The assembly of multilayer structures out of these single-layers is a promising direction to combine the physical
properties of them for the design of a new generation of optical devices.
Different stackings of 2D materials lead to different band alignments.
This allows to design the charge transfer properties upon optical
excitation by choosing suitable 2D heterostructures.\cite{Grigorieva2013,Frisenda2018,Yankowitz2012,Yu2013,Lee2014,Furchi2014,Cheng2014,Withers2015,Ursula2017} 

Multilayer systems offer the possibility for the formation of interlayer excitons besides the
intralayer ones in single-layer
2Ds.\cite{Gong2013,Kang2013,Terrones2013,Kosmider2013,Komsa2013,Torun2016,Yagmurcukardes2016,Miller2017,Galvani2016}
This makes TMDs based heterobilayers (HBLs) potential candidates for ultrafast charge
transfer,\cite{Hong2014} ultrafast formation of hot interlayer excitons,\cite{Chen2016}
interlayer energy transfer,\cite{Kozawa2016}, valleytronics,\cite{Schaibley2016,Molina-Sanchez2017}
charge transfer,\cite{Fang2014,Gong2014,Tongay2014,Zhu2015,Rigosi2015,Zhang2016,Hill2016,Furchi2018}, and 
long-lived interlayer excitons \cite{Rivera2015}. In addition, recent efforts have described the role of 
the Moir\'{e} patterns in TMD HBLs on the binding energy of excitons.\cite{Wu2018}

On the theoretical side, efforts have focused on the electronic structure, predicting the
type-II alignment for several stacking combinations of TMDs using density functional theory (DFT) calculations. 
\cite{Gong2013,Kang2013,Terrones2013,Kosmider2013,Komsa2013,Torun2016,Yagmurcukardes2016,Zibouche} 
For the compounds with type-II band alignment, on the independent particle level, the interlayer transition is the lowest energy transition 
due to the band offset of the constituent single layers. However, excitonic
effects might reverse the order of intra versus inter-layer transitions.

The reason is that the spectral position of the interlayer excitons with respect to intralayer ones 
depends not only on the band alignment but also on the excitonic binding energy which is strongly enhanced in 2D materials (as compared to bulk materials). The exciton binding energy depends on the distance between charge carriers via the screened Coulomb interaction. As the electron and hole of the interlayer excitons are spatially separated, the binding energy of them is, a priori, weaker than that of intralayer ones. At the same time, however, it is known (for bulk layered systems) that the screening in the perpendicular direction is weaker than the screening in the layer-plane, which, in turn, tends to enhance the binding energy of inter-layer excitons. If the binding energy of the interlayer exciton
is much smaller than the lowest intralayer one, the optical band alignment of the compound might deviate from the electronic band alignment. 
Therefore, the understanding of the optical response and the carrier dynamics of TMD HBLs requires an accurate calculation of the excitonic states together with the GW correction of the electronic structure.

In this work, we report the electronic structure and
optical absorption spectra, including excitonic effects and full spinorial wave
functions,
of MoS$_2$/WS$_2$ and MoSe$_2$/WSe$_2$ HBLs. We classify the intralayer
and interlayer excitons and report the valence and conduction band
alignments. We find that the lowest energy exciton of both HBLs has interlayer character
(charge transfer state), which makes these systems suitable to host
excitons with long lifetimes. We find good agreement with the spectral ordering of excitonic peaks in
the recent photoluminescence measurements of the MoSe$_2$/WSe$_2$ bilayer by Wilson et al.\cite{Wilson2017}

\section{Methods}
We calculate the excitonic states and the optical absorption spectra of MoS$_2$/WS$_2$ 
and MoSe$_2$/WSe$_2$ HBLs using \textit{ab initio} many-body perturbation theory with the
Bethe-Salpeter Equation (BSE).\cite{Strinati1982,Rohlfing2000,Palummo2004} In this formalism, the
excitations are expressed in terms of electron-hole pairs:
\begin{equation}
  (E_{c\textbf{k}}-E_{v\textbf{k}})A^{S}_{vc\textbf{k}}+ 
  \sum_{\textbf{k}'v'c'}
  \langle vc \textbf{k} |  K_{eh} | v'c'\textbf{k}'
  \rangle A^{S}_{v'c'\textbf{k}'} =
  \Omega^{S} A^{S}_{vc\textbf{k}}
\end{equation}
where $E_{c\textbf{k}}$ and $E_{v\textbf{k}}$ are the quasi-particle energies of
the valence and the conduction band states, respectively. The energies and
wave functions are obtained from DFT as implemented in
Quantum Espresso\cite{QE} using the local density approximation (LDA) and norm-conserving fully
relativistic pseudo-potentials.\cite{Hamann2013} The pseudopotentials
are generated based on the parameters of PseudoDojo \cite{Setten2018}.
The plane wave energy cutoff is 120 Ry.
We use fully relativistic pseudopotentials,
Mo and W semi-core electrons are included in the calculations. The vacuum distance between two periodic 
images is 40 a.u. for both the single- and bi-layers. 
In order to get the quasi-particle eigenvalues, the LDA energies are 
corrected by the G$_0$W$_0$ approximation,\cite{Hedin1970,Onida2002} as implemented in the Yambo
code.\cite{yambo}
The G$_0$W$_0$
quasi-particles energies are calculated on a $42 \times 42 \times  1$ $\mathbf{k}$-grid, centered on $\Gamma$.
We use 160 bands for the self-energy and 160 bands for the dynamical dielectric
screening. 

The $A^{S}_{vc\textbf{k}}$ are the expansion
coefficients of the excitonic states, and $\Omega^{S}$ are their energies. The
 interaction kernel between electrons and holes, $K_{eh}$,
contains the unscreened exchange interaction $V$ (repulsive) and the screened direct Coulomb
interaction (attractive) $W$. The latter term $W$ depends on the
dielectric screening. In the case of 2D materials the accurate treatment of
the dielectric screening is crucial. The lower dielectric screening (when compared with 3D materials)
results in large exciton binding energies, of the
order of 0.5 eV.\cite{Molina-Sanchez2013,Molina-Sanchez2015,Qiu2013}

The imaginary part of the dielectric function, $\epsilon(\hbar \omega)=
\epsilon_{1}(\hbar \omega)+ i \epsilon_{2}(\hbar \omega)$, is proportional to
the optical absorption spectra. It is expressed, in terms of the excitonic states as
\begin{equation}
\epsilon_{2}(\hbar \omega) \propto
\sum _{S}
\Bigg|  \sum _{cv\textbf{k}} A^{S}_{vc\textbf{k}}
        \dfrac{\langle c \textbf{k} | p_{i} | v \textbf{k} \rangle}
        {(\epsilon_{c \textbf{k}}-\epsilon_{v \textbf{k}})}
\Bigg|^{2}
\delta(\Omega^{S}-\hbar \omega)
\label{osc_strength}
\end{equation}
where $\langle c \textbf{k} | p_{i} | v \textbf{k} \rangle$ are the dipole
matrix elements of transitions from the valence to the conduction bands. We
consider in-plane polarization for both single and bilayers. The out of plane
absorption gives a negligible contribution at the bandgap energies due to
depolarization effects. In order to mimic the experimental results, the delta
function is replaced by a Lorentzian with $0.05$ eV broadening. Similar to G$_0$W$_0$, 
the BSE calculation are also performed using the Yambo code.\cite{yambo} In order to avoid the longe-range 
interaction between the periodic copies of the single-layer along the vertical direction, a Coulomb cutoff 
of the screened potential is used in both in G$_0$W$_0$ and BSE calculations. 
Since we are dealing only with the low energy part of the absorption spectra, it
is sufficient to include only the 4 highest valence bands and 4 lowest conduction
bands in the Bethe-Salpeter kernel.

\section{Results}
The lattice parameters of MoS$_2$ (MoSe$_2$) and WS$_2$ (WSe$_2$) single-layers
are almost commensurate\cite{Komsa2013,Yu2015} which justifies the construction of the HBLs assuming
A-A$^{'}$ stacking (see the geometries in Fig. \ref{fig1}(a)) with the
bulk lattice parameters of
3.162 \AA{} and 3.288 \AA{} for MoS$_2$/WS$_2$\ and
MoSe$_2$/WSe$_2$ HBLs, respectively.\cite{2Dbook} In our ground
state calculations, we use these
experimental lattice parameters without performing further optimization but we have relaxed
the atomic positions and the distance between layers on the LDA level. Even though the LDA completely neglects the 
Van der Waals interaction between layers, it gives reasonable interlayer distances for many layered systems because it overestimates the weak covalent contribution 
to the interlayer bonding. \cite{Molina-Sanchez2015} 

\subsection{Orbital-projected band structure}
\begin{figure}
  \includegraphics[scale=0.6]{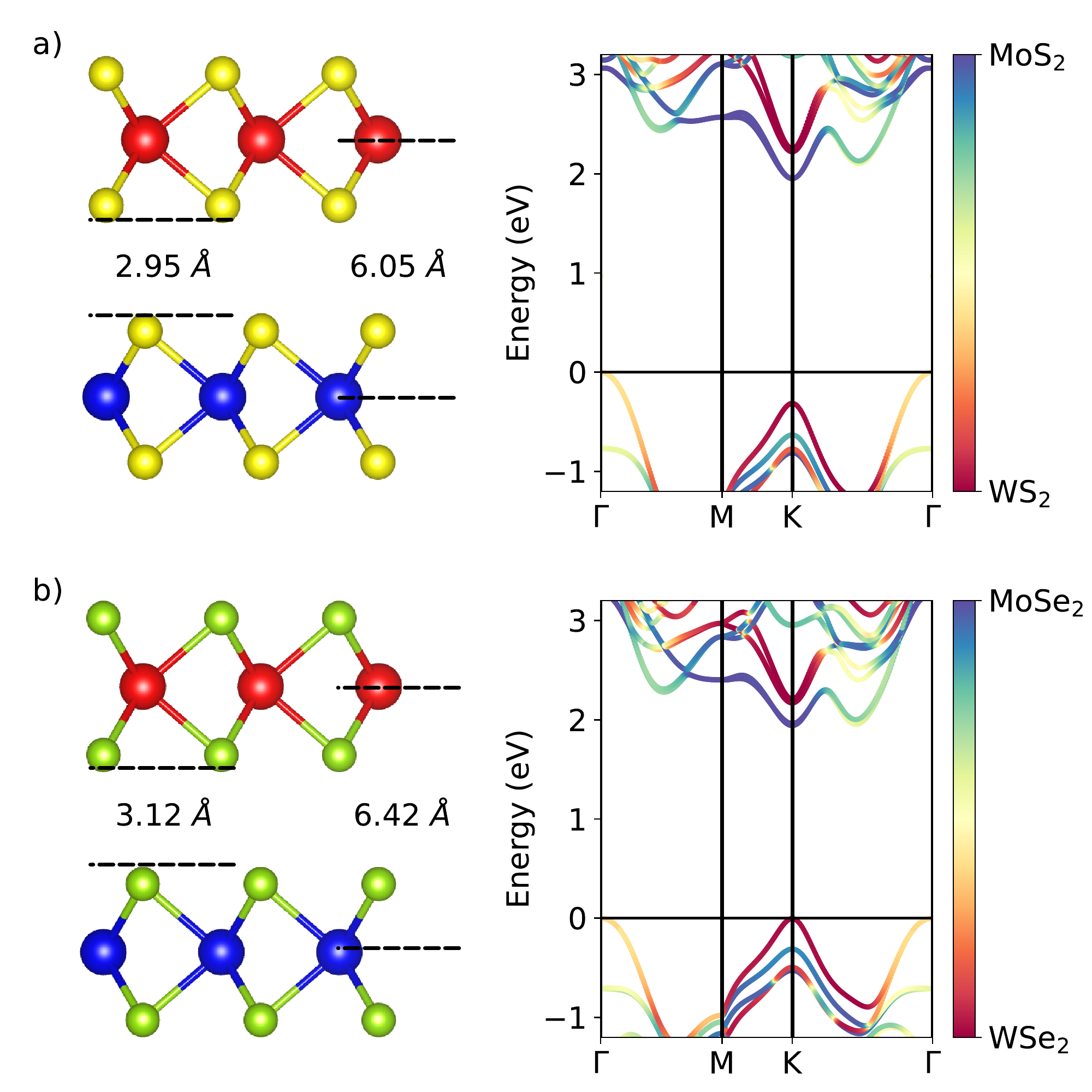}
  \caption{Optimized atomic and projected electronic structures of (a)
  MoS$_2$/WS$_2$ and (b) MoSe$_2$/WSe$_2$ HBLs.
  The red, blue, yellow and light green atoms correspond to W, Mo, S
  and Se, respectively.}
  \label{fig1}
\end{figure}

Even without calculating the optical properties, the electronic structure of the HBLs already offers
valuable information. Figures \ref{fig1}(a) and (b) show the band structures of MoS$_2$/WS$_2$\ and MoSe$_2$/WSe$_2$ HBLs, respectively.
The bands in the figure are projected on to the atomic orbitals of the single layers. Therefore, bands with blue and red color correspond to
MoS$_2$ (MoSe$_2$) and WS$_2$ (WSe$_2$) layers, respectively.

As can be seen in Fig.\ref{fig1}, the conduction and valence bands at the $K$ point in the Brillouin Zone (BZ) 
are not hybridized and can be assigned unambiguously to the constituent single-layers. The
trend is that the conduction band minima are purely localized on the MoX$_2$ whereas the
valence band maxima are localized on the WX$_2$ layers where X represents S and Se atoms. 
Therefore, on the LDA level, the band alignment of HBLs are type-II, with conduction band (electrons) and valence
band (holes) located at different layers. The G$_0$W$_0$ calculations changes the
magnitude of the alignments but not the character. Moreover, the
order of the valence band states is determined by including properly the
spin-orbit interaction. In our calculations the spin-orbit interaction is included exactly using full
spinor wave functions. Table \ref{table1} reports the values of bandgaps, band offsets and spin-orbit splitting, as obtained in
LDA and G$_0$W$_0$ levels. These conclusions are in line with the experimental observations. \cite{Wilson2017}

\begin{table}
\caption{\label{table1} Direct LDA and G$_{0}$W$_{0}$ energy gap, SOC splitting
and band offset at the point K in the BZ of the  MoS$_2$/WS$_2$ and
MoSe$_2$/WSe$_2$ HBLs and constitute single-layers. VB and CB stands for valence and conduction band, respectively. The band offset values in G$_{0}$W$_{0}$ are shown in parantheses.}
\begin{tabular}{ccccccccc}
\hline\hline
                        &     \multicolumn{2}{c}{Energy Gap          }&  SOC splitting & \multicolumn{2}{c}{Band offset}\\
                        &     \multicolumn{2}{c}{       (eV)         }&      (eV)      & \multicolumn{2}{c}{   (eV)    }\\
                        &     LDA         &     G$_{0}$W$_{0}$        &                &   VB                &      CB          \\
\hline
   MoS$_2$              &    1.62         &       2.53                &     0.16       &  ---                &     ---          \\
    WS$_2$              &    1.56         &       2.52                &     0.46       &  ---                &     ---          \\
MoS$_2$/WS$_2$          &    1.29         &       2.26                &      ---       &  0.32(0.22)         &    0.27(0.23)   \\
  MoSe$_2$              &    1.38         &       2.19                &     0.20       &  ---                &    ---           \\
   WSe$_2$              &    1.30         &       2.23                &     0.49       &  ---                &    ---           \\
MoSe$_2$/WSe$_2$        &    1.07         &       1.95                &      ---       &  0.31(0.24)         &     0.21(0.19)  \\
\hline\hline
\end{tabular}
\end{table}

In addition, the interlayer interaction makes the HBLs indirect semiconductors
on both the LDA and the G$_0$W$_0$ level. As shown in Fig.\ref{fig1}, the
valence band maximum is located at $\Gamma$ and composed of hybrid orbitals from
both layers. The conduction band minimum is located at the $K$ point and
composed of non-hybridized Mo orbitals.
Our calculations also show that MoS$_2$/WS$_2$ and MoSe$_2$/WSe$_2$ are
indirect bandgap semiconductors with a LDA (G$_0$W$_0$) gap of 1.10 (1.92)
and 1.05 (1.74) eV which is consistent with the
results for hetero-bilayer systems.\cite{Kosmider2013}

\subsection{Optical spectra}
In layered compounds, excitonic effects are much stronger than in bulk compounds due to reduced Coulomb 
screening. The prominent excitonic effects are particularly important for the low energy optical response and the 
charge carrier dynamics of the ultra thin materials. In the case of HBLs, the excitons posses inter or intralayer character whose 
spectral position depend on their binding energy and the band alignment of the constituent single-layers. 
Therefore, the type-II band alignment of the HBLs obtained in the independent-particle picture can be
insufficient to ensure an interlayer exciton at the lowest energy in the optical spectra. 
Thus, a realistic calculation of excitonic binding energies and a characterization of the optical properties of TMD HBLs demands for accurate \textit{ab initio} methods with the BSE approach including the spin-orbit coupling.

\begin{figure*}
\includegraphics[scale=0.60]{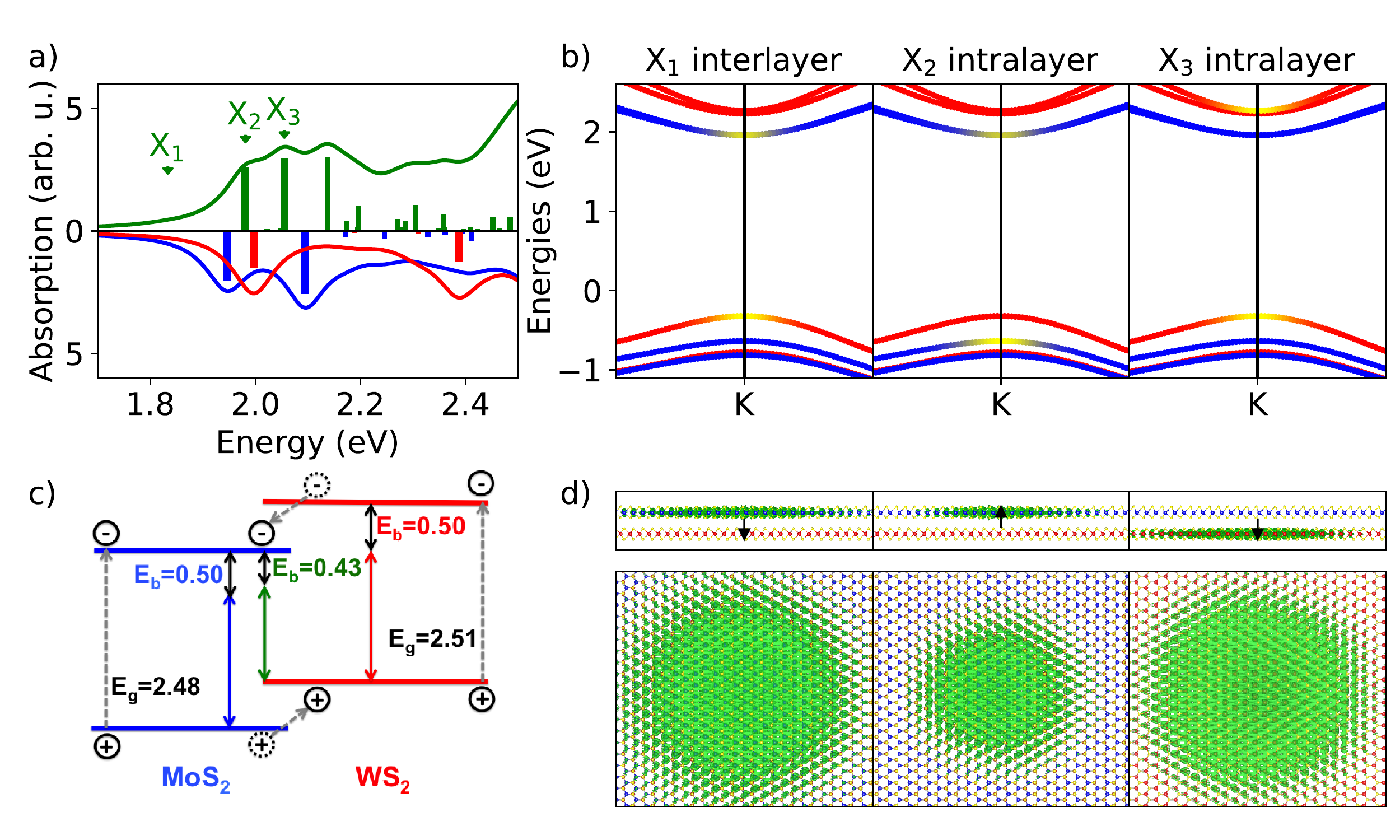}
\caption{Optical absorption spectra of (a) MoS$_2$/WS$_2$ and constituent single-layers
(blue and red curves for Mo and W compound, respectively).
(b) Electronic bands near the K point in the BZ with the transitions
contributing to the exciton. (c) band alignment of the HBLs with excitonic effects.
(d) The charge density of the indicated excitons with a fixed hole position
marked with a black arrow.}
\label{fig2}
\end{figure*}

\begin{figure*}
\includegraphics[scale=0.60]{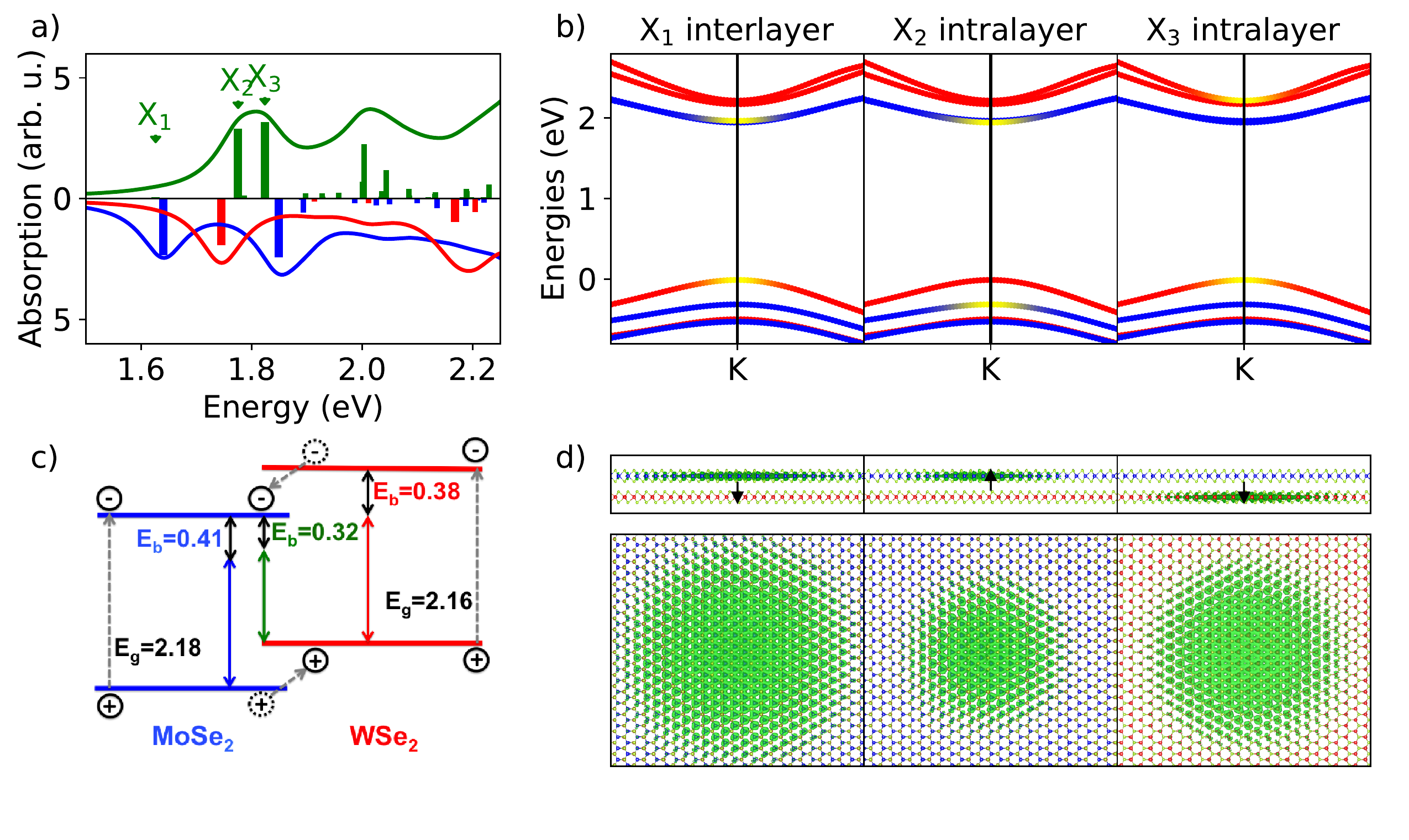}
\caption{Optical absorption spectra of (a) MoSe$_2$/WSe$_2$ and constituent single-layers
(blue and red curves for Mo and W compound, respectively).
(b) Electronic bands near the K point in the BZ with the transitions
contributing to the exciton. (c) band alignment of the HBLs with excitonic effects.
(d) The charge density of the indicated excitons with a fixed hole position
marked with a black arrow.}
\label{fig3}
\end{figure*}

The optical spectra including excitonic effects of MoS$_2$/WS$_2$ and
MoSe$_2$/WSe$_2$ HBLs are shown in Figures \ref{fig2} and \ref{fig3},
respectively. In both figures, the panel (a) shows the absorption
spectra for the constituent layers MoX$_2$ (blue), WX$_2$ (red)
and HBLs (green). We focus on the absorption threshold of the spectra, in
particular on the first three bright excitons of each HBLs.

In the case of MoS$_2$/WS$_2$ HBL the $X_1$ exciton is an interlayer exciton 
which is energetically lower than the intralayer ones as shown in Fig. \ref{fig2}(a). 
The projected band structure (Fig. \ref{fig2}(b)) shows that the exciton is composed of transitions
from the of top of the highest valence band at K to the minimum of the second conduction band (note that the
spin-orbit splitting of the conduction band minimum is only 
3 meV \cite{Echeverry2016} and thus the two lowest conduction bands cannot be
distinguished on the energy scale of Fig. \ref{fig2}). 
The exciton wave function is represented by fixing the hole and plotting the
electron density. In all the figures, the maximum of the electron density is set to 1
and we fix a consistent isosurface value. The wave function of the exciton localizes in the WS$_2$ layer when
the hole is placed in the MoS$_2$ layer as can be see in Fig. \ref{fig2}(d). \footnote{In order to diagonalize the
Bethe-Salpeter Hamiltonian and obtain the wave functions of the first excitonic states we
implemented a new diagonalization mode in the Yambo code using the SLEPc library
(http://slepc.upv.es/). This library provides iterative algorithms to obtain the
N lowest energy eigenvalues and corresponding eigenvectors in a selected energy range.}
The small oscillator strength of this exciton peak is the result of the
spatially separated charge carriers. Another important
point is that the interlayer $X_1$ exciton is not the lowest energy exciton in the absorption spectrum of the HBL, there is another 
interlayer dark exciton $D_1$, which is 3 meV lower in energy (and due to transitions from the top valence to the lowest conduction band). Therefore, photoluminescence (PL) can be quenched at
low temperatures if the splitting dark-bright is large enough (as noted already for PL from intra-layer excitons\cite{Zhang2015,Echeverry2016}).

In addition to the interlayer exciton we also present the first two intralayer excitons
derived from band-to-band transitions within each single-layer. It can be seen in the projected band structure plot that the intralayer 
excitons X$_2$ and X$_3$ belong to MoS$_2$ and WS$_2$ layers, respectively. The localization of the electron and the hole in the same layer (see Fig. \ref{fig2}(d))
enhances the oscillator strength and therefore the absorption is much stronger
than the interlayer exciton. These excitons are slightly red-shifted with
respect to the single-layer excitons (see blue and red spectra in Fig.~
\ref{fig2}(a)), as a result of the increased dielectric screening in the 
case of bilayer.

\begin{table*}
\caption{\label{table2}The spectral position, composition and
binding energy of the excitons indicated in Fig. \ref{fig1}. We include the
peak positions of the $X_2$ and $X_3$ excitons in the case of single-layer.}
\begin{tabular}{c|cccc|cccc}
\hline\hline
                                         &          \multicolumn{4}{c|}{MoS$_2$/WS$2$}                            &    \multicolumn{4}{c}{MoSe$_2$/WSe$2$}   \\
\hline
&    D$_1$  &     X$_1$   &    X$_2$    &     X$_3$ &  D$_1$  &  X$_1$   &    X$_2$    &     X$_3$     \\
\hline
\multirow{2}{*}{Spectral position (eV)}           & \multirow{2}{*}{1.830}
&\multirow{2}{*}{1.833} & \multirow{2}{*}{1.98 (1.95)}   & \multirow{2}{*}{2.05
(1.99)}& \multirow{2}{*}{1.60} & \multirow{2}{*}{1.63}  & \multirow{2}{*}{1.77
(1.64)}        & \multirow{2}{*}{1.82 (1.75)}  \\
\multirow{2}{*}{Binding energy (eV)}             & \multirow{2}{*}{0.43} &\multirow{2}{*}{0.43} & \multirow{2}{*}{0.50}   & \multirow{2}{*}{0.50} & \multirow{2}{*}{0.32}& \multirow{2}{*}{0.32}  & \multirow{2}{*}{0.41}        & \multirow{2}{*}{0.38}  \\
\multirow{2}{*}{Composition}             & \multirow{2}{*}{W-Mo} &\multirow{2}{*}{W-Mo} & \multirow{2}{*}{Mo-Mo}  & \multirow{2}{*}{W-W}  & \multirow{2}{*}{W-Mo}  &\multirow{2}{*}{W-Mo}  & \multirow{2}{*}{Mo-Mo}       & \multirow{2}{*}{W-W}  \\
\multirow{2}{*}{           }             & \multirow{2}{*}{ } &\multirow{2}{*}{ } &\multirow{2}{*}{ } & \multirow{2}{*}{ }  & \multirow{2}{*}{ } & \multirow{2}{*}{ }  & \multirow{2}{*}{ }        & \multirow{2}{*}{ }  \\
\hline
\hline
\end{tabular}
\end{table*}
The excitonic binding energies of the HBLs provide valuable information
of their optical properties. Table \ref{table2} shows that the 
interlayer exciton, X$_1$, of MoS$_2$/WS$_2$ HBL has a binding energy of 0.43 eV which is 70 meV smaller
than that of the first intralayer exciton, X$_2$, originating from MoS$_2$ layer. 
This is an expected outcome since the charge carriers of interlayer excitons are spatially separated which reduces the binding energy.
Yet, the weaker Coulomb screening between layers prevent the binding energy of interlayer exciton from being much smaller than the binding energy of 
the intralayer ones. The competition of these two contributions together with the large band offset ultimately leads to a sufficiently large binding energy such that
the interlayer exciton is the lowest energy one. Therefore, the optical properties of a MoS$_2$/WS$_2$  HBL 
correspond to the ones of a type-II heterostructure in spite of the strong
excitonic effects of 2D materials. Note, however, that the difference in excitonic effects reduces the
energy difference between inter and intralayer exciton to 150 meV as opposed to
an energy difference of 0.260 that would be expected in the independent particle model (neglecting excitonic binding energy).
It is worth to mention that high accuracy
of first principles calculations is required to obtain a reliable result. The omission
of the spin-orbit interaction (up to 0.5 eV for WS$_2$) and/or of the
Coulomb cutoff can dramatically change the result and the conclusions.

Regarding the MoSe$_2$/WSe$_2$ HBL, the analysis of Fig. \ref{fig3} gives
similar physical conclusions but significant quantitative differences. The
binding energies are smaller than that of the previous case, i.e. the binding energies of interlayer
and intralayer excitons are 0.32 eV and 0.41 eV, respectively. Similar to the previous case, 
the MoSe$_2$/WSe$_2$ HBL displace the character of type-II band alignment, both on the independent-particle 
level and when excitonic effects are included.   
In addition, the dark exciton (30 meV below the bright one!) can be be clearly distinguished from 
the band structure shown in Fig. \ref{fig3}(b) for exciton $X_1$.
We expect the effect of PL quenching to be more visible at low
temperature for MoSe$_2$/WSe$_2$ HBL than in the case of MoS$_2$/WS$_2$. 

\begin{figure}
  \includegraphics[scale=0.7]{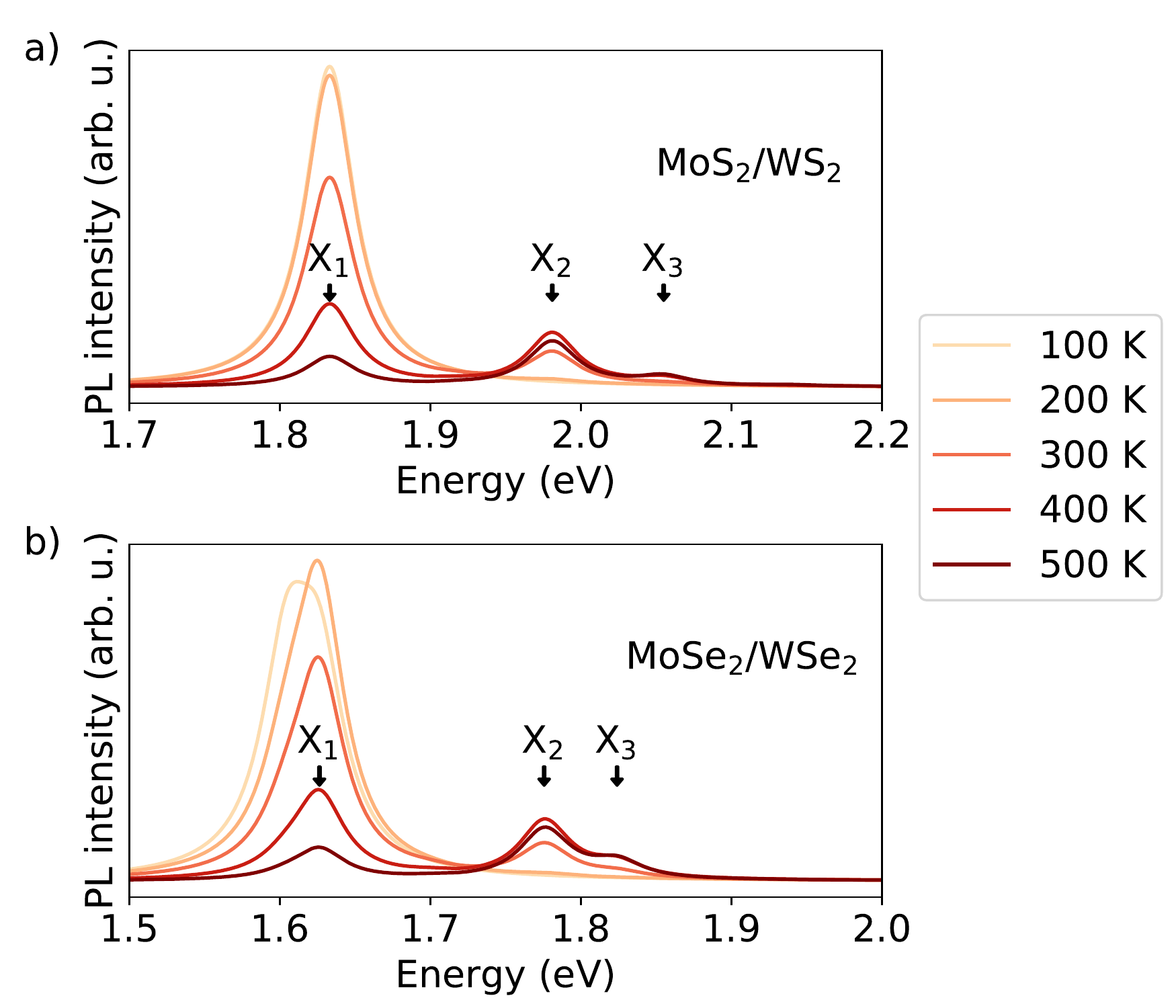}
  \caption{Visualization of the photoluminescence intensity for the
MoS$_2$/WS$_2$ and
MoSe$_2$/WSe$_2$ systems in panel a) and b) respectively obtained by multiplying the oscillator strength by the
Bose-Einstein distribution $(e^{E_s/{k_bT}}-1)^{-1}$ for different temperatures
where $E_s$ is the energy of the exciton. }
  \label{fig4}
\end{figure}
Experimental proofs of the existence of interlayer excitons are more robust for the Se HBL than for the
S HBL case. In photoluminescence experiments, e.g., Wilson et. al.\cite{Wilson2017} detected the 
intra-layer exciton peaks at 1.57 and 1.64 eV for MoSe$_2$ and WSe$_2$,
respectively (in qualitative agreement with our
results in Table~\ref{table2}) and the interlayer exciton around 0.22 eV below
the $X_2$, in comparison with our prediction of 0.14 eV. Differences can be due to
the presence of a substrate, which is neglected in our calculations and due to the bilayer twist or stacking that results from the layer depositions in the experiments. Time-dependent
PL showed long lifetime excitons with low radiative efficiency, indicating
that the lowest energy exciton has interlayer character in agreement with our results.\cite{Zhang2015}
We present only calculations of absorption spectra where the intensity of the peaks is directly given
by the dipole matrix elements (oscillator strengths) of the excitonic states (see Eq.~\ref{osc_strength}).
For the Se HBL, the oscillator strength of the interlayer exciton is fifty
times smaller than the one of the lowest intralayer exciton (due to the spatial
separation of the wave functions on neighboring layers).
As can be seen in Fig 3 (a), the intralayer exciton can hardly be detected in an
absorption experiment.
In PL experiments, however, the intensity ratio 
of the peaks is reversed. The intensity is proportional to the oscillator strength 
and to the exciton population. Since the exciton recombination time is slower
than the thermalization\cite{Rivera2015,Palummo2015}, we assume on a first
approximation that the occupation of the
excitonic states follows the Bose-Einstein distribution. 
Figure~\ref{fig4} shows a visualization the PL intensity for several
temperatures, obtained by multiplying
the oscillator strength by the Bose-Einstein distribution.
The ratio between two peaks is then proportional to the Boltzmann 
factor $\exp(-\Delta E/k_BT)$, where $\Delta
E$ is the energy difference between two excitons.
An improved quantitative calculation of luminescence spectra would include the
transition rates from the intralayer excitons (into which absorption takes
place) to the interlayer exciton. However, these rates are currently still
unknown.
A formal theoretical treatment of the PL can be found in Ref.~\onlinecite{Melo2016}
but is beyond the scope of the present work.

\section{Conclusions}
Our first-principle investigation on the excitons of MoS$_2$/WS$_2$ and
MoSe$_2$/WSe$_2$ HBLs predict the existence of interlayer excitons, 0.15 eV and
0.24 eV
below the absorption onset of intralayer excitons. This indicates that the excitonic ground
states of these systems naturally separate the electron and the hole in
different layers, making TMDs HBLs efficient materials for charge separation
applications. We also observe that the lowest energy exciton of both HBLs is dark with a remarkable splitting of 30 meV 
with respect to first bright interlayer exciton for MoSe$_2$/WSe$_2$ bilayer. Our calculations 
agree well with available experimental data\cite{Rivera2015,Wilson2017} within the limits imposed
by the uncertainties about heterostructure geometry (e.g., twisting angle) and influence of
the substrate via screening effects. We also obtain good agreement 
with recently reported calculations for the MoSe$_2$/WSe$_2$ HBL.\cite{Gao2017,Gillen2018}
The dipole oscillator strength of the interlayer exciton is 50 times smaller than the one of the
intralayer excitons. This means that in the absorption spectra, the corresponding peak is practically invisible. In luminescence spectra, it becomes the dominant peak. The quantitative description of the measured luminescence spectra at room temperature\cite{Wilson2017} and at 20 K\cite{Rivera2015} requires the calculation of transition rates from the intra- to the interlayer excitonic states which is the subject of future work.

\section{Acknowledgement}
The authors acknowledge support by the National Research Fund, Luxembourg, through the projects
INTER/ANR/13/20/NANOTMD (E.T and L.W.), C14/MS/773152/FAST-2DMAT (A.M.S.), and OTPMD (H.P.C.M.). 
Furthermore, A. M.S acknowledges the Juan de la Cierva MEC program. H.P.C.M acknowledges support 
from the F.R.S.-FNRS through the PDR Grants HTBaSE (T.1071.15).
The simulations were done using the HPC facilities of the University of Luxembourg~\cite{VBCG_HPCS14}.
We acknowledge stimulating discussion with Jens Kunstmann who raised the
question of the relatively large binding energy of the inter-layer exciton
during the presentation of our work at the Flatlands 2017 conference in Lausanne.

\bibliographystyle{apsrev-nourl}
\bibliography{./biblio.bib}

\end{document}